\documentclass[conference,10pt]{IEEEtran}
\usepackage{cite}
\usepackage[dvips]{graphicx}
\usepackage{graphicx}
\usepackage{color}
\usepackage{amsmath}
\usepackage{amstext}
\usepackage{hyperref}
\usepackage{amsfonts}
\usepackage{amssymb}
\usepackage{psfrag}
\usepackage{fancyhdr}

\usepackage{subfigure}
\usepackage{epsfig}
\usepackage{csquotes}
\MakeOuterQuote{"}

\usepackage[ruled,vlined]{algorithm2e}

\usepackage{tabularx}

\usepackage{etoolbox}
\let\bbordermatrix\bordermatrix
\patchcmd{\bbordermatrix}{8.75}{4.75}{}{}
\patchcmd{\bbordermatrix}{\left(}{\left[}{}{}
\patchcmd{\bbordermatrix}{\right)}{\right]}{}{}

\begin{document}

\title{Exploring Social Networks for Optimized User Association in Wireless Small Cell Networks with Device-to-Device Communications}

\author{
\IEEEauthorblockN{Muhammad Ikram Ashraf\IEEEauthorrefmark{1}, Mehdi Bennis\IEEEauthorrefmark{1}, Walid Saad\IEEEauthorrefmark{2} and Marcos Katz\IEEEauthorrefmark{1} \\}
\IEEEauthorblockA{\small\IEEEauthorrefmark{1}Centre for Wireless Communications, University of Oulu, Finland, \\ email: \{ikram,bennis,marcos.katz\}@ee.oulu.fi \\
\IEEEauthorrefmark{2}Electrical and Computer Engineering Department, University of Miami, Coral Gables, FL, USA, email: walid@miami.edu}
%\vspace{-.5cm}
%\thanks{This work was supported by the Finnish Funding Agency for Technology and Innovation (Tekes) under the %Celtic-Plus "Green Terminals for Next Generation Wireless Systems" (Green-T CP8-006) %project}
}

\maketitle

\begin{abstract}
In this paper, we propose a novel social network aware approach for user association in wireless small cell networks. The proposed approach exploits social relationships between user equipments (UEs) and their physical proximity to optimize the network throughput. We formulate the problem as a matching game between UEs and their serving nodes (SNs). In our proposed game, the serving node can be a small cell base station (SCBS) or an \emph{important node} with device-to-device capabilities. In this game, the SCBSs and UEs maximize their respective utility functions capturing both the spatial and social structures of the network. We show that the proposed game belongs to the class of matching games with \emph{externalities}. Subsequently, we propose a distributed algorithm using which the SCBSs and UEs interact and reach a stable matching. We show the convergence of the proposed algorithm and study the properties of the resulting matching. Simulation results show that the proposed socially-aware user association approach can efficiently offload traffic  while yielding a significant gain reaching up to $63\%$ in terms of data rates as compared to the classical (social-unaware) approach.

% Check again Abstract based on Mehdi's comments.
% ...........*fill in metric used here*.............

%We simulate the problem as a matching game which take into account individual and distributed decisions on whether to cooperate or nor, while maximizing a utility function that captures the cooperative gains, in terms of rate and social ties. We show that preference used by every node in the network is independent and influenced by the network. Due to the nature of individual and distributed decisions at every node in the network, we classified as a many-to-one matching game with externaliities.
%The simulation results shows that the proposed social-aware approach outperforms the classical approach by up to $45\%$ with the significant increase in gains as compared to the social un-aware approach.

\end{abstract}

\IEEEpeerreviewmaketitle

\section{Introduction}
\label{sec:intro}
%\textbf{INTRODUCTION NEED TO BE FIXED}

The proliferation of bandwidth intensive wireless applications such as mobile video streaming and social networking has led to a tremendous data demand increase \cite{TQS2013}. This increasing need for wireless capacity mandates novel cellular architectures to deliver high quality-of-service (QoS) while maintaining a cost-effective operation. In this respect, small cell networks, built on the premise of densely deploying small cell base stations (SCBSs) has emerged as a promising technique. Small cell networks revolve around the idea of shrinking cell sizes and, thus, bringing users closer to their base stations so as to boost the network capacity. Reaping the promised performance benefits from small cell deployments requires overcoming many fundamental challenges that includes user association, network modeling, interference mitigation, and resource management, among others \cite{JA2013, TQS2013, DLopez2011, GdeRoche2010}. Beyond small cell networks, the introduction of device-to-device communications overlaid on macro or small cell networks has recently emerged as a promising technique that can further improve the performance of wireless networks.
%Remember to add References.
One key challenge in D2D-enabled small cell networks is the problem of associating user equipments (UEs) to their preferred serving node (SN) that can be a small cell or other D2D users. Classical user association mechanisms in small cell networks are based on physical layer aspects e.g., signal-to-interference-plus-noise ratio (SINR), signal strength, or delay and do not account for the presence of D2D \cite{JA2013, TQS2013, DLopez2011, GdeRoche2010, RMadan2010}. In \cite{RLD2d2013, GED2d2012,KMD2d2009}, the authors studied D2D approach that exploits the physical and wireless attributes to improve the resource allocation, throughput, spectrum and energy efficiency of the underlaid cellular network. While such classical approaches are suitable for traditional services such as voice, the emergence of smartphones and social networking necessitate leveraging new dimensions, going beyond traditional physical layer parameters. The possibility of exploiting these new dimensions coupled with the heterogeneity of small cells networks render the user association problem challenging. Recently, in \cite{YL2013}, the authors study a D2D-overlaid network exploring the social network and physical layer characteristics to improve the network performance and reduce the load. Therein, exploring the underlying \emph{social ties} among users allow operators to better cope with peak data traffic demands. However, in \cite{YL2013}, the social ties among nodes depend only on the probability that a certain content is selected and the proposed algorithm for data sharing is heuristic.

The main contribution of this paper is to exploit notions of social networks and underlaid D2D links to optimize user association in small cell networks. In particular, we utilize social popularity metrics and we introduce the concept of \textit{important UE} which acts as the best SN for other UEs within the D2D proximity range. In this respect, we explore the \emph{social ties} among UEs to identify the set of important UEs in the network. Then, by identifying socially important UEs within a geographical area, we leverage the D2D links to offload the traffic load of the small cell network, thereby enhancing the user association procedure. Using the proposed model, the decisions of the UEs on whether to use a cellular or D2D link takes into account the social importance of the node as well as the network load, channel condition and interference. We formulate the problem as a social-aware matching game with externalities in which the serving nodes SNs (i.e., SCBS, important UE) and UEs are the players that rank one another so as to find a suitable and stable association. Here, UEs rank one another and the SCBSs via a utility function that captures both the social and spatial structure of the network.

We show that the proposed game belongs to a class of matching games with \emph{externalities} due to the interference and interdependence among preferences. These externalities relate to the unique features of small cell networks such as interference, as well as and the underlying social features of the networks. This is in contrast to existing works on matching for wireless such as in \cite{EA2011, SBayat2012, ALeshem2011} which ignore interference and externalities. To solve the studied game, we propose a distributed algorithm that allows the UEs and SNs to interact so as to optimize their utilities (i.e., social welfare) that leads to a stable matching. Simulation results validate the effectiveness of the proposed approach and show significant performance gains compared to the baseline social-unaware user association approach.

The rest of this paper is organized as follows. In Section \ref{sec:2}, we present the system model and we describe the studied social network model. In Section \ref{sec:3}, we formulate the social aware UE-SN association problem as a matching game with externalities, and we propose an algorithm to solve it. Simulation results are presented in Section \ref{sec:4}. Finally, conclusions are drawn in Section \ref{sec:5}.

\section{System Model}
\label{sec:2}

\subsection{Wireless Network Model}
Consider the \textit{downlink} transmission of a single carrier macrocell network. A number of small cell base stations (SCBSs) are overlaid on this macrocell network. Let $\mathcal{N} = \{1,..., N\}$, $\mathcal{M} = \{1, ..., M \}$, and $\mathcal{I} = \{1,..., I\}$ where $\mathcal{I} \subset M,\; \mathcal{I} \neq \mathcal{M}$, represent the set of SCBSs, the set of all UEs, and the set of all \emph{important UEs}, respectively. An important UE is a socially popular node in the social network which is overlaid on the wireless network. We let $\mathcal{P}$ be the set of $P$ SNs which can be either SCBSs or \emph{important UEs} $\mathcal{P} \subseteq \mathcal{N} \bigcup \mathcal{I}$. In classical cellular systems, each UE is typically serviced by the SCBS with the highest received signal strength indicator (RSSI). Let us suppose that $\mathcal{L}_{i}$ is the set of UEs serviced by SCBS $i$. We assume that the Bandwidth $BW_i$ of each SCBS $i$ is divided equally between the served UEs. Here, we assume that each UE $m$ will utilize a single subcarrier $c$. The achievable rate between SCBS $i$ and UE $m$ over subcarrier $c$ is:
%In order to serve each UE $m \in \mathcal{L}_{i}$,
\begin{align}
\label{eq:capacity}
			R_{i,m} = BW_{i} \cdot \log_2 \left(1+ \frac{p_{i,m,c} h_{i,m,c}}{\sigma^2+\sum_{j \in \mathcal{N} \setminus \{i\}} p_{j,m,c} h_{j,m,c}} \right),
\end{align}
%\hat{I}_{i,m}
\noindent
where $h_{i,m,c}$ is the channel gain over subcarrier $c$ used by UE $m$ to transmit to SCBS $i$ and $\sigma^2$ is the variance of the Gaussian noise. The term summation represents the aggregate interference at UE $m$ caused by the transmissions of other SCBSs on the same subcarrier $c$.

%$h_{i,m}$, $BW_{i,m}$ are the channel gain and subcarrier between UE $m$ and SCBS $i$ respectively, and $\sigma^2$ is the variance of the Gaussian noise. The term $\hat{I}_{i,m} = \sum p_{j,m} h_{j,m}, j \in \mathcal{N} \setminus \{i\}$ represents the aggregate interference at UE $m$ caused by the transmissions of other SCBSs on the same subcarrier $BW_{i,m}$.

\subsection{Social Network Model}
We study the social connections between the users present within the coverage of the considered cellular network. In particular, these social networks can be represented by graphs in which the vertices represent the users and the edges represent the relationships between the users based on friendship, common interests and use of shared services, among others. Here, we let $G_s = (\mathcal{V},\mathcal{E})$ denote a social graph, where $\mathcal{V}$ defines as the set of all $V$ nodes and $\mathcal{E} \subseteq \mathcal{V} \times \mathcal{V}$ is the set of edges. Adjacent nodes $(m,n)$ are connected via a bidirectional edge $e \in \mathcal{E}$, where an edge $e$ between adjacent nodes reflects SCBS-UE or UE-UE links.

\subsection{Important UE}
An UE which is socially popular as compared to other UEs in the network is considered as an important UE. The three most popular ways to quantify the social popularity of nodes in the social graph are degree centrality, closeness, and betweenness \cite{DrWhite1994, PC2006}. The social importance of an UE can be defined by exploring the concept of the edge betweenness centrality, similarity, and physical distance to the peer node in the social network.

\textbf{Edge betweenness Centrality:}
Edge betweenness centrality is based on the idea that an edge becomes central to a graph if it lies between many other nodes, i.e., it is traversed by many of the shortest paths connecting a pair of nodes \cite{PC2006}. Edges with a high betweenness centrality are considered important because they control information flow in the social network. Let $\boldsymbol{B} = \{b_{m,n}\}$ be the edge betweenness matrix, where $b_{m,n}$ is the edge betweenness of the link between nodes $m$ and $n$. The betweenness centrality $b_{m,n}$ of an edge $e$ between nodes $(m,n)$ is the sum of the fraction of all-pairs shortest paths that pass through edge $e$ \cite{Ulrik2008}. The normalized $b_{m,n}$ is:% defined as:
\begin{equation}
\label{eq:betweenness}
	%\begin{split}
			 b_{m,n} = \frac{ \displaystyle \sum_{m,n \in \mathcal{V}} \frac{\rho (m,n|e)}{\rho(m,n)} } {(V-1) (V-1)},
			 %b_{m,n} & =  \frac{b_{m,n}}{(V-1)(V-2)},
	%\end{split}
\end{equation}
\noindent
where $V$ is the number of nodes, the summation over the number of shortest $(m,n)$-paths, and $\rho (m,n|e)$ is the number of those paths that traverse edge $e$.

\textbf{Similarity Matrix:}
The similarity is a measure of closeness between a pair of nodes. The degree of similarity can be measured by the ratio of common neighbors between individuals in a social networks. The degree of similarity between nodes $m$ and $n$ has an important effect in terms of data dissemination. The nodes having lower degree of similarity are good candidate for data dissemination. Hence, for a pair of nodes $(m,n)$, depending on whether they are connected directly or indirectly, their similarity matrix $\boldsymbol{Q}$ is defined \cite{TZhou2009}:% as \cite{TZhou2009}:
\begin{equation}
\label{eq:similarity}
			q_{m,n} = \begin{cases}	\displaystyle \sum_{z \in \nu(m)\cap\nu(n)}\frac{1}{k(z)} & \mbox{if $m,n$ are connected},\\
        $0$ & \mbox{otherwise}.\end{cases}
\end{equation}
Let $\boldsymbol{Q}$ is a $V \times V$ matrix, $\nu(m)$ is the set of neighbors of $m$, $\;z \in \nu(m) \cap \nu (n)$ are the common neighbors of nodes $m$ and $n$, and $k(z)$ is the degree of UE $z$. To normalize the similarity matrix, we use the simple additive weighting (SAW) method. Thus, the normalized value of each element $q_{m,n}$ of $\boldsymbol{Q}$ is:
\begin{equation}
\label{eq:normsim}
	s_{m,n} = q_{m,n}/q_{n}^{\max} \;\; \forall m,n,
\end{equation}
where $q_{n}^{\max}=\max_{m} q_{m,n}$. Consequently, we obtain the normalized similarity matrix $\boldsymbol{S}=\{s_{m,n}\}$, where element $s_{m,n}$ denotes the normalized similarity between nodes $m$ and $n$.

\textbf{Social Distance:}
%The social distance is defined as the social interaction parameter between communicating nodes. Selecting the UE that provides the maximum utility for data transfer is referred to as \textit{important UE}. Let $X$ be a matrix where element $x_{m,n}\in[0,1]$ quantifies how the social distance of a user affects its utility which denotes the social distance parameter between two nodes $(m,n)$ \cite{Eli2007}:
The social distance is defined as the social interaction parameter between communicating nodes. The \textit{important UEs} in a given cell are defined as the subset of UEs with the highest social distance for data transfer. Let $\boldsymbol{X}$ be a matrix where element $x_{m,n}\in[0,1]$ quantifies how the social distance of a user affects its utility which denotes the social distance parameter between two nodes $(m,n)$ \cite{Eli2007}:
%\begin{equation}
\begin{align}
	\label{eq:socialdistance}
	 \boldsymbol{X} = \alpha \boldsymbol{S} + \beta \boldsymbol{B},
\end{align}
%\end{equation}
where $\alpha$ and $\beta$ are tunable parameters such that $\alpha + \beta = 1$.

The \textit{important node} selects its preferred peer based on the composite social and physical distance captured by the following weighted cost function $w$.
\begin{equation}
			\label{eq:weightcost}
			w_{p,m} = \epsilon_{p,m} d_{p,m} {x}_{p,m}
\end{equation}
We combine the social distance with actual physical distance between UEs, where $x_{p,m}$ is the social distance, $d_{p,m}$ is the physical distance between nodes $p$ and $m$ and $\epsilon_{p,m}$ is a normalization constant. In order to illustrate how the important UE is selected in the network, we presented the following example.

\begin{figure}[t]
\centering
\includegraphics[scale = 0.45]{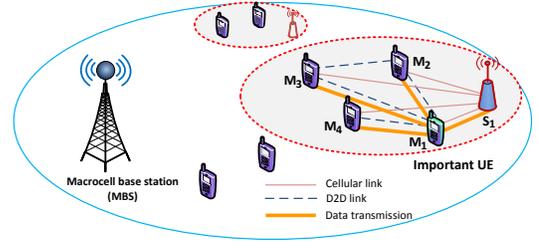}
\caption{Illustrative example of the considered network deployment.}
\label{fig:network-model}
\end{figure}

\textbf{Example:}
Consider four UEs, within the set $\mathcal{M} = \{M_{1},M_{2},M_{3},M_{4}\}$ and one SCBS, $S_{1}\in \mathcal{N}$ as shown in Fig. \ref{fig:network-model}. Mathematically, the connectivity of node $m\in \mathcal{M}$ can be represented by an adjacency matrix $\boldsymbol{A}$, which is a $V \times V$ symmetric matrix, where $V$ is the number of nodes in the social graph $G_s$. The adjacency matrix has elements:
\begin{equation}
			A_{m,n} = 	\begin{cases}	1 & \mbox{if there is a edge between nodes $m$ and $n$},\\
        0 & \mbox{otherwise}.\end{cases}
				\nonumber
\end{equation}
%\vspace{3 mm}
The edge betweenness centrality between nodes is given by $\boldsymbol{B} = \{b_{m,n}\}$, where each element of $b_{m,n}$ denotes the edge betweenness among two nodes $m$ and $n$.

$
\boldsymbol{B} = \bbordermatrix{~ &  &  &  & \cr
     	S_{1} & 0     & 0.083 & 0.125 & 0.125 & 0.167 \cr
			M_{1} & 0.083 &   0   & 0.125 & 0.125 & 0.167 \cr
		  M_{2} & 0.125 & 0.125 & 0     & 0.083 & 0     \cr
		  M_{3} & 0.125 & 0.125 & 0.083 & 0     & 0     \cr
      M_{4} & 0.167 & 0.167 & 0     & 0     & 0     \cr}
$
\vspace{2 mm}
\noindent

For a social graph $G_s$ with $V$ nodes, the similarity between each pair of nodes is computed using (\ref{eq:similarity}). The normalized similarity matrix $\boldsymbol{S}$ between node pair $(m,n)$ is given by:

$
\boldsymbol{S} = \bbordermatrix{~ &  &  &  & \cr
     	S_{1} & 0     &   1   & 0.583 & 0.583 & 0.25 \cr
			M_{1} & 1     &   0   & 0.583 & 0.583 & 0.25 \cr
		  M_{2} & 0.583 & 0.583 & 0     & 0.50  & 0.50 \cr
		  M_{3} & 0.583 & 0.583 & 0.50  & 0     & 0.50 \cr
      M_{4} & 0.25  & 0.25  & 0.50  & 0.50  & 0    \cr}
$

$
\boldsymbol{X} = \bbordermatrix{~ &  &  &  & \cr
  S_{1} &       0  & 0.5208 &  0.3227 &  0.3227  & 0.1667 \cr
  M_{1} &  0.5208  &      0 & 0.3227  &  0.3227  & 0.1667 \cr
  M_{2} &  0.3227  & 0.3227 &      0  & 0.2708   & 0.2500 \cr
  M_{3} &  0.3227  & 0.3227 & 0.2708  &      0   & 0.2500 \cr
  M_{4} &  0.1667  & 0.1667 &  0.2500 & 0.2500   & 0      \cr}
$
\vspace{2 mm}
To determine which UE is the most important node for the SCBS, consider $\alpha = 0.5 $ and $\beta = 0.5$. By looking at the social distance $\boldsymbol{X}$ matrix, it is clear that UE $M_{1}$ is the most socially important than other UEs while, $M_{4}$ is the least important one. Hereinafter in this work we consider one important UE per small cell.

%\textbf{Remark:} Note that, node preference is considered in the first place. The assignment is to optimize utility with the condition that every node preference is respected. Our paper is the first application of the social network graph in cell association in wireless communications to the best of our knowledge.

\section{Social network aware user association as a matching game with Externalities}
\label{sec:3}

%\subsection{Problem Formulation}
To solve the proposed user association model with social awareness, we will use the framework of matching theory \cite{Eliz2011}. Prior to defining the game, we will propose socially-aware utility functions using which UEs and SCBSs can assess the performance they achieve during transmission, given both the wireless and social network metrics.

\subsection{Utility of UE}
UE $m$ can either connect to any SCBS $i$ or instead connect to an \textit{important UE} $p$, via a \textit{D2D link}. If UE $m$ prefers to be matched with the \textit{important node} instead of SCBS then the achievable rate of UE $m$ is $\frac {\min(R_{i,p}, R_{p,m})} {2} $. The utility $U_{i,m}$ captures the tradeoff between the achievable rate among SCBS $i \in \mathcal{N}$ and UE $m \in \mathcal{M}$ for the given matching $\eta$, expressed as:
\begin{equation}
\label{eq:utility-ue}
			U_{i,m} (R_{i,m},x_{i,m}, \eta) = \begin{cases}	\frac{R_{i,m}}{x_{i,m}} & \mbox{if $m=p$},\\
        R_{i,m} & \mbox{if $m$ connected to SCBS $i$}, \\
        \frac {\min(R_{i,p}, R_{p,m})} {2}  & \mbox{for D2D}.
        \end{cases}
\end{equation}
\subsection{Socially-Aware Utility of SCBS}
To calculate the utility of SCBS $i \in \mathcal{N}$, we incorporate the social distance of each UE $m$ with respect to SCBS $i$ \cite{Strat2011}. For the given matching $\eta$, the utility of an SCBS $i$ is the sum of utilities of its associated UEs $m \in \mathcal{L}_{i}$. Thus, the utility of SCBS $i\in \mathcal{N}$ is:
\begin{align}
\label{eq:utility-scbs}
	U_{i}(\eta) & = \displaystyle \sum_{m \in \mathcal{L}_{i}} U_{i,m}(R_{i,m}, x_{i,m}, \eta).
%	U_{i}(R, X, \eta) & = \displaystyle \sum_{m \in \mathcal{I}_{i}} U_{m}(R, X, \eta) & \notag \\
%	&  =	\displaystyle \sum_{m \in \mathcal{I}_i} U_m(R_{i,m}) \cdot X_{i,m}
\end{align}

The proposed user association approach leverages the social and spatial structure of the D2D-underlaid small cell network.  In order to avoid the combinatorial complexity nature of the problem, we make use of the framework of matching games \cite{Eliz2011}. To associate UEs to SNs, each SN aims at identifying the largest set of UEs, for which it can meet the minimum social distance requirements. Having defined such utilities, our goal is to solve the problem of associating each UE $m \in \mathcal{M}$ to the best serving node $k \in \mathcal{P}$ via a matching $\eta: \mathcal{M} \rightarrow \mathcal{P}$. In what follows, we formulate the user association problem as a social matching game, and discuss its properties.

A \textit{social matching game} is defined by two sets of the players $\mathcal{M}$, the set of UEs and $\mathcal{P}$, the set of SNs and two preference relations $\succ_m$, $\succ_j$ for each UE $m \in \mathcal{M}$ (SN $j \in \mathcal{P}$) to build preference over SN $j \in \mathcal{P}$ (UE $m \in \mathcal{M}$) and vice-versa. A \emph{preference relation} $\succ_m$ is defined for every UE $m$ over the set of SNs $\mathcal{P}$ such that for any two nodes in $j , k \in \mathcal{P}, j \neq k $ and two matchings $\eta, \eta\prime \in \mathcal{M} \times \mathcal{P}, \eta
\neq \eta\prime\;$:
\begin{align}
    & (j,\eta) \succ_m (k, \eta\prime) \Leftrightarrow & \notag \\
    &U_{j,m}(R_{j,m}, x_{j,m}, \eta) > U_{k,m}(R_{k,m},x_{k,m}, \eta\prime),
%    &U_{j,m}(R_{j,m}, X_{j,m}, \eta) > U_{k,m}(R_{k,m}, X_{k,m}, \eta\prime),
\end{align}
where $(j,m) \in \eta$ and $(k,m) \in \eta\prime$. Similarly the preference relation $\succ_j$ for the SN $j$ over the set of UEs $\mathcal{M}$ is defined as, for any two UEs $m,n \in \mathcal{M}, m \neq n$
\begin{equation}
(m,\eta) \succ_j (n, \eta\prime) \Leftrightarrow U_j(\eta) > U_j(\eta\prime)
\end{equation}

Each SN and UE independently rank one another based on the respective utilities in (\ref{eq:utility-ue}) and (\ref{eq:utility-scbs}) that capture the social ties among UEs. The selection preferences of UEs are interdependent and influenced by the existing network wide matching, which leads to a many-to-one matching game with externalities.

% Re-write the paragraph.
\subsection{Social Welfare}
%We aim to solve the problem of assigning each UE $m \in \mathcal{M}$\ to the best serving node (SN) $k \in \mathcal{P} $ via a matching $\eta : \mathcal{M} \rightarrow \mathcal{P}$.
Our goal is to solve the problem of matching many UEs to the best SNs (many-to-one). In this matching, many UEs are matched to each SN, and UEs have preferences over the SNs. The social welfare $W$ \cite{Eliz2011} is expressed as the sum of the utilities of UEs and SCBSs.
\begin{align}
\label{eq:socialwelfare}
    W(\eta) = \displaystyle \sum_{k \in \mathcal{P}} U_{k}(\eta) + \displaystyle \sum_{m \in \mathcal{M}} U_{k,m} (R_{k,m}, x_{k,m}, \eta)
\end{align}
\noindent
The social-aware user association problem can be written as:
\begin{eqnarray}
\label{eq:matching}
& \displaystyle \max_{\eta:(k,m) \in \eta} W(\eta), \\
& s.t,\;\;\;\;\; R_{k,m} \geq \bar{R}_{k,m},\; \forall \; m
%D_{i,m} \geq \overline{D}_{i,m}
\end{eqnarray}
\noindent
where $\bar{R}_{k,m}$ is the minimum target rate per UE. In terms of complexity, solving the UE-SN association problem using classical optimization techniques is an NP-hard problem \cite{NPHard2006}, which depends on the number of SNs and UEs in the network. In this work, we propose a heuristic procedure which is based on two-way preferences for both UEs and SNs. Specifically, the proposed approach leverages the spatial and social structure of the networks, in which UEs social relationships are taken into account for establishing suitable D2D connections.
\begin{algorithm}[t]
\footnotesize
\caption{Socially aware user association algorithm}
\label{algo:algo1}
\DontPrintSemicolon % Some LaTeX compilers require you to use \dontprintsemicolon instead
\KwData{Each UE is initially associated to a randomly selected SCBS $i$.}
\KwResult{Convergence to a stable matching $\eta$.}

\textbf{Phase I - SN discovery and utility computation;}

\begin{itemize}
	\item Each UE $m$ discovers a serving node (SN) in the vicinity;
    \item UEs and SNs exchange social-aware information $X$ and achievable rate metrics in (\ref{eq:capacity}), (\ref{eq:socialdistance});
%	\item UEs and SNs exchange the social data $\boldsymbol{X}$ (\ref{eq:socialdistance}) and rate $R$ (\ref{eq:capacity});
	\item We have important UEs list $\mathcal{I}$ using (\ref{eq:socialdistance});
	\item $U_{j,m}(\eta)$ and $U_{k}(\eta)$ and social welfare $W(\eta)$ are updated;

\end{itemize}
\textbf{Phase II - Swap-matching evaluation;}
    \begin{itemize}
	   \item Pick the random pair of UEs $\{m,n\} \in \mathcal{M}$;
	\end{itemize}
	\While{$count \leq \max$ \textit{Iterations}} {
	\begin{itemize}
			 \item $U_{j,m}(\eta)$, $U_{k}(\eta)$ are updated based on the current $\eta$;
				\item UEs and SNs are sorted by $W(\eta)$;
				\item swap the pair of UEs $\{m,n\} \{n,m\}$
				\item $W(\eta)_{n,m} = W_{best}(\eta)$
	\end{itemize}
     $P_T = \frac{1}{1+e^{-T (W(\eta)_{n,m} - W(\eta)_{m,n})}}$;\\
     $\eta_{m,n} \leftarrow \eta_{n,m}$ change the configuration with probability $P_T$;\\
		    \If{$W(\eta)_{n,m} > W(\eta)_{best}$}{
					$W_{best} = W(\eta)_{n,m}$\;
    }
    \Else{
      SN $k$ refuses the proposal, and UE $m$ sends a proposal to the next configuration at $count$ \;
    }
		$count = count + 1$
}
\textbf{Phase III - Social-aware user association;}
\end{algorithm}

In the proposed algorithm (Algorithm \ref{algo:algo1}), the preferences carried out by both the UEs and SNs are done locally, whereas coordination is required between adjacent SCBSs. If a UE $m \in \mathcal{M}$ is not currently served by its most preferred SN $k \in \mathcal{P}$, it sends a matching proposal to another SN $j$. Upon receiving a proposal, SN $j$ updates its utility and accepts the request of the UE if the externalities resulting from such swap do not yield a degradation of the social welfare. The main goal of each UE is to maximize its own utility while associating with the most preferred SCBS or important UE. Our game is a matching game with \emph{externalities} due to SINR and social-ties between nodes, which is a class of games that significantly differs from existing works such as \cite{EA2011, SBayat2012, ALeshem2011}. To deal with the externalities arising due to interference and coupling among different matchings, we allow a two-sided swapping between pairs of SCBS and UEs, as a function of the prospective increase in social welfare. The concept of externalities leads to a new stability concept based on the idea of a stable swap-matching \cite{Eliz2011}:

\textbf{Definition 1:} Given a matching $\eta$, a pair of UEs $m,n\in\mathcal{M}$
and SNs $j,k \in \mathcal{P}$ with $\eta_{m,n} = \{\eta \{(m,j),(n,k)\} \cup \{(m,k),(n,j)\}\}$ a swap-matching is \emph{stable} if:
$ \forall y \in \{m,n,j,k\}$, such that $U_{y}(\eta_{k,j}) \leq U_{y}(\eta_{j,k}) $ and $\nexists y \in \{m,n,j,k\}\;U_{y}(\eta_{k,j}) > U_{y}(\eta_{j,k}) $\\
 	
A swap-matching $\eta$ is said to be stable with link $(j,m) \in \eta$ if there does not exist any UE $n$ or SN $k$, for which SN $j$ prefers UE $n$ over UE $m$. Such network-wide matching stability is reached by guaranteeing only stable swap-matching among players \cite{Eliz2011} which result in an increase in the social welfare. Classical matching approaches are based on preference orders, such as the deferred acceptance algorithm \cite{ALeshem2011}. Nevertheless, such approaches do not account for externalities, and thus may yield lower utilities or may not converge for a game with externalities. In fact, due to externalities, players continuously change their preference orders, in response to the formation of other UE-SN links. Inspired from \cite{Eliz2011}, we use a Markov Chain Monte Carlo (MCMC) algorithm. Hence, instead of using the greedy way to select the ``best'' matching, the matching is chosen based on a probability, which depends on the swap which results an increase of social welfare $W$.

\begin{figure}[t]%
\centering%
\includegraphics[scale = 0.40]{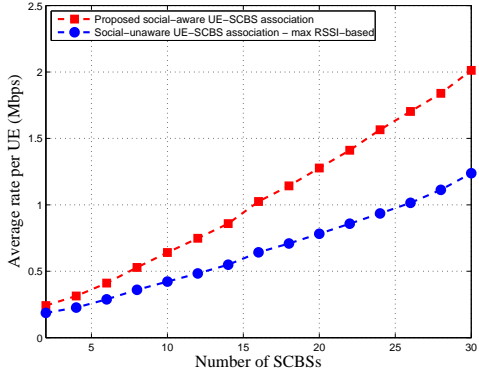}
\caption{Average rate per UE as a function of the number SCBSs $N$ for fixed number of UEs $M = 150$, under the considered approaches.}
\label{fig:avg-rate}
\end{figure}%

For finding a stable matching in the proposed small cell user association, we use Algorithm \ref{algo:algo1}. The value of temperature $T$ goes to zero as the iteration $count$ approaches the $Tth$ iteration step \cite{Temp2012}. In the first phase, each UE is initially associated to a randomly selected SN based on the max-RSSI criterion. Next, each SCBS gets the social information from the UEs to identify the important nodes. Then, the social welfare is calculated for the current matching $\eta$. In the second phase, UEs and SNs update their respective utilities and individual preferences over one another. Subsequently, at each iteration, a chosen UE pair is swapped with a probability that depends on the change in the social welfare: a positive change in the social welfare yields a probability of swapping larger then 1/2 and vice-versa. As a result, the algorithm does not get caught in a local optimum. The algorithm continuously keeps track of the
``best'' matchings found so far \cite{Eliz2011}.

The proposed algorithm is guaranteed to converge. The convergence of Algorithm \ref{algo:algo1} follows from two facts. First, since only the swaps which verify the conditions of Definition 1 are allowed, the utility of each player cannot decrease after each swap as each swap guarantees non-decreasing utilities which result into an increase of the social welfare $W$. Second, the utility of SCBS is the sum of the utilities of UEs connected to it whose number is finite. Moreover, the utility of the \emph{important UE} is the sum of the utilities of the UEs connected to it. If Algorithm 1 is run sufficiently long it can find the two-sided stable matching. However, there is no guarantee that the best matching encountered in finite time is even two-sided stable, a situation that can be remedied by applying greedy approaches. Algorithm \ref{algo:algo1} terminates when no further improvement can be achieved. Therefore after Phase II the algorithm converges to a local maximum of the social welfare W. Finally, uniform power allocation is used.

\section{Simulation Results}
\label{sec:4}
For simulations, we consider a single micro-cell in which UEs and SCBSs are uniformly distributed over the area of interest. Transmissions are affected by distance dependent path loss according to 3GPP specifications \cite{3GPP}, while the tunable variables $\alpha$ and $\beta$ are set to $0.5$. For the simulation we considered there is no power control, where the power is uniformly divided between the UEs. The simulation parameters are given in Table \ref{tab:simulation}. We compare the performance of the social un-aware approach in which UEs associate based on the max-RSSI criterion.

%For simulations, we consider a single micro-cell with a radius of $500m$, SCBS coverage radius of $50$m and the D2D transmission radius of $20$ m. SCBSs and UEs are uniformly distributed over the area with transmit power of $23dbm$ and $15dbm$ respectively. The noise power spectral density is $-174$ dbm and the path loss for D2D communication is $(d)^{-\alpha}$ $\alpha=3$. Transmissions are affected by distance dependent path loss according to 3GPP specifications \cite{3GPP}, while the tunable variables $\alpha$ and $\beta$ are set to $0.5$. For the simulation we considered there is no power control, where the power is uniformly divided between the UEs. We compare the performance of the social un-aware approach in which UEs associate based on the max-RSSI criterion.

Fig. \ref{fig:avg-rate} shows the average rate per UE as a function of the SCBSs $N$, with $M=150$. Fig. \ref{fig:avg-rate} highlights the fact that, in the proposed social-aware approach, user associations result in enhanced SINRs and increasing transmission rates. Fig. \ref{fig:avg-rate} shows that, as the number of SCBSs increase, the average rate increases. This is due to the fact that, an increase in the number of SCBSs decreases the density of UEs per SCBS. In Fig. \ref{fig:avg-rate}, we can also see that the proposed social-aware matching game yields significant performance gains, increasing with $30\%$, and reaching up to $63\%$ when the number of SCBS $N=30$ relative to the social unaware association approach.

\begin{table}[b]
\caption{Simulation Parameters}
\centering
\begin{tabular}{| p{5.5cm} | p{2.5cm} |}
\hline
\textbf{Parameter} & \textbf{Value} \\ \hline
%\hline
System bandwidth                                        & 5 MHz \\ \hline
Macro cell radius                                       & 500 m \\ \hline
Noise power spectral density ($N_0$)                    & -174 dBm/Hz \cite{3GPP} \\ \hline
SCBS transmission radius                                & 50 m \\ \hline
SCBS transmission power                                 & 23 dBm \\ \hline
D2D transmission radius                                 & 20 m \\ \hline
UE transmission power                                   & 15 dBm \\ \hline
Device-to-Device (D2D) pathloss  & $(d)^{-\alpha}$, $\alpha = 3$\\ \hline
\end{tabular}
\label{tab:simulation}
\end{table}

\begin{figure}[t]%
\centering%
\includegraphics[scale = 0.46]{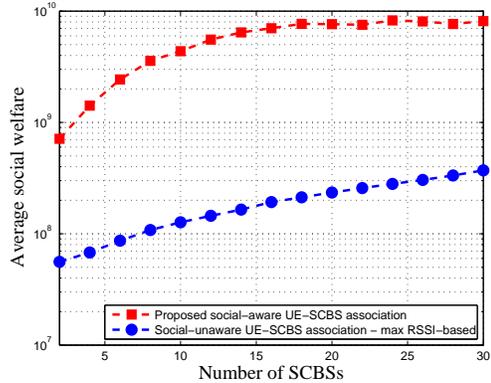}
\caption{Average social welfare for the fixed number of UEs $M = 150$, under the considered approaches.}
\label{fig:welfare-plot}
\end{figure}%

Fig. \ref{fig:welfare-plot} shows the average social welfare for fixed $M = 150$ UEs. In this figure, we can see that the increase in the number of SCBSs result in an increase in the average social welfare of about $21\%$ at $N=30$, relative to the social unaware approach. The social welfare is the sum of the utilities of the UEs and SCBSs, and, thus by adding more UEs and SCBSs Fig. \ref{fig:welfare-plot} shows an increase in average social welfare.

\begin{figure}[t]%
\centering%
\includegraphics[scale = 0.44]{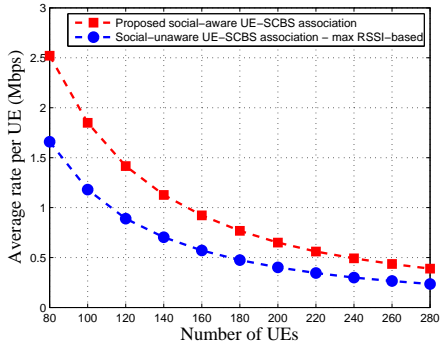}
\caption{Average rate per UE as a function of the number of UEs $M$ for a fixed number of SCBSs $N = 16$, under the considered approaches.}
\label{fig:vary-ue-rate}
\end{figure}

Fig. \ref{fig:vary-ue-rate} shows the average rate per UE as a function of the UEs $M$ for a fixed number of SCBSs $N = 16$. The decrease in the average data rate is due to the increased interference as number of UEs increases in the system. Fig. \ref{fig:vary-ue-rate} shows that the proposed approach yields significant gains, in terms of the average rate per user, reaching up to $65\%$ when the UEs $M=280$ compared to the max RSSI scheme.

\begin{figure}[t]%
\centering%
\includegraphics[scale = 0.44]{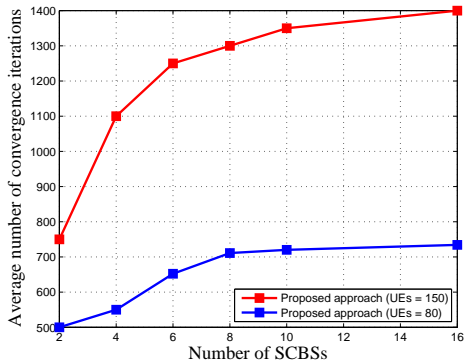}
\caption{Average number of iterations as a function of the number of the SCBSs $N$ and fixed number of UEs, under the proposed approach.}
\label{fig:avg-iterations}
\end{figure}

Fig. \ref{fig:avg-iterations} shows the average number of iterations required to reach a stable matching $\eta$ as a function of a fixed number of UEs $M=80$ and $M=150$ and varying number of SCBSs $N$. In this figure, we can see that for lower number of SCBSs, the number of iterations needed to converge increases with increasing number of SCBSs and reaches up to $730$, $1400$ iteration at SCBSs $N=16$ for fixed UEs $M=80$ and $M=150$ respectively. However, a further increase in the number of SCBSs for a fixed number of UEs results in lower number of UEs per SCBS. Consequently, an increase in the number of SCBSs $N$ decreases the number of possible D2D pairs and, thus, will not lead to a high increase in the number of iterations. Moreover, for a fixed number of SCBSs, higher iterations are incured as the number of UEs per small cells increases.

\vspace{-0.3cm}
\section{Conclusions}
\label{sec:5}
In this paper, we have presented a novel, social-aware approach to the user association problem in D2D underlaid small cell networks. We have formulated the problem as a matching game with externalities in which the UEs and SNs build preferences over one another so as to choose their own utilities. Simulation results have shown that the proposed social-aware approach provides considerable gains in terms of increased data rates with respect to a traditional social-unaware UE-SN association which is based on maximum RSSI. In our future work, we will further incorporate issues of power control and arbitrary number of important users.


\section*{Acknowledgment}
This work is supported by the Finnish Funding Agency for Technology and Innovation (Tekes) under the Celtic-Plus Green Terminals for Next Generation Wireless Systems (Green-T CP8-006) project, as well as
U.S. National Science Foundation under Grant CNS-1253731.

%%%%%%%%%%%%%%%%%%%%%%%%%%%%%%%%%%%%%%%%%%%%%%%%%%%%%%%%%%%%%%%%%%%%%%%%%%%%%%%%%%%%%%%%%%%%%%%%%%%%%%%%%%%%%%%%%%%%%%%%%%%%%%%%%%%%%%%%%%%%%%


%\bibliographystyle{IEEEtran}
%\bibliography{IEEEabrv,IEEEbibl}

\renewcommand{\baselinestretch}{0.85}

\end{document}